\documentclass[prb,showpacs,twocolumn]{revtex4}

\usepackage{amsmath,amssymb}

\usepackage{graphicx}

\begin{document}

\widetext

\title{Peierls Transition in the Quantum Spin-Peierls Model}

\author{William Barford$^{1}$ and Robert J. Bursill$^2$}

\affiliation{
$^1$Department of Physics and Astronomy, University of Sheffield, Sheffield, S3 7RH, United Kingdom\\
$^2$School of Physics, University of New South Wales, Sydney, New South Wales 2052, Australia
}

\begin{abstract}
We use the density matrix renormalization group method  to investigate the role of longitudinal quantized phonons on the Peierls transition in the spin-Peierls model. For both the XY and Heisenberg spin-Peierls model we show that the staggered phonon order parameter scales as $\sqrt{\lambda}$ (and the dimerized bond order  scales as $\lambda$) as $\lambda \rightarrow 0$ (where $\lambda$ is the electron-phonon interaction). This result is true for both linear and cyclic chains. Thus, we conclude that the Peierls transition occurs at $\lambda=0$ in these models. Moreover, for the XY spin-Peierls model we show that the quantum predictions for the bond order  follow the classical prediction as a function of inverse chain size for small $\lambda$. We therefore conclude that the zero $\lambda$ phase transition is of the mean-field type.
\end{abstract}

\pacs{75.10.Jm, 71.38.-k}

\maketitle


Since the 1930s the role of electron-phonon interactions on the stability of the metallic state in one-dimensional metals has been of abiding interest to both theoretical chemists and physicists. In the 1950s Fr\"ohlich \cite{frohlich} and Peierls \cite{peierls} showed that as a consequence of electron-phonon interactions a one-dimensional metal is unstable with respect to an insulating (or Peierls) ground state. Equivalent conclusions were later derived for conjugated polymers by Ooshika \cite{ooshika}, and Longuet-Higgins and Salem \cite{longuet}. All of these results were derived in the adiabatic limit (or Born-Oppenheimer approximation) in which the phonon degrees of freedom are regarded as the slow, classical variables. In this limit an infinite metallic system undergoes a Peierls transition to the lower symmetry insulating ground state for any value of the electron-phonon interaction.

The question of whether such a conclusion remains valid for quantized phonons was first addressed by Fradkin and Hirsch \cite{fradkin}. Using both renormalization group arguments and Monte Carlo simulations they suggested that quantum fluctuations do not destroy the Peierls state for non-interacting electrons. However, they also suggested that quantum fluctuations can destroy the Peierls state for non-interacting spinless fermions, giving a non-zero value for the critical electron-phonon interaction.

More recent numerical calculations on various models have also indicated that the Peierls state may be destroyed by quantized phonons. Caron and Moukouri \cite{caron} used finite size scaling analysis of the spin gap to show that at a critical value of the electron-phonon interaction there is a Kosterlitz-Thouless transition in the XY spin-Peierls model with  \textit{gapped, non-dispersive} (or Einstein) phonons. Similar conclusions were made for the Heisenberg spin-Peierls model, again with Einstein phonons\cite{bursill99, uhrig}. Finally, the Holstein model with Einstein phonons for both spinless\cite{bursill98} and spinfull\cite{jeckelmann} fermions, 
and the extended Hubbard-Peierls model with Einstein \emph{bonds} phonons\cite{sengupta}
have also be shown to exhibit a Peierls phase transition at a non-zero value of the electron-phonon interaction. 

In this article we focus our attention on the role of quantized \textit{gapless, dispersive} (or Debye) phonons on the Peierls  state. We investigate both the XY spin-Peierls and Heisenberg spin-Peierls models. Spin-Peierls models are the strong coupling limit of the Hubbard-Peierls model at half-filling. For open, linear chains the XY spin-Peierls model is equivalent to the non-interacting spinless fermion-Peierls model, allowing us to make direct comparisons to the essentially exact adiabatic predictions. For both models we find that although quantized phonons reduce the amplitude of the broken symmetry order parameter, in contrast to the previous investigations, there is no evidence that the Peierls state is destroyed at a finite value of the electron-phonon interaction.

The Heisenberg spin-Peierls model with dispersive, gapless phonons is defined by,
\begin{eqnarray}\label{Eq:1}
    H =  H_1 + H_2
\end{eqnarray}
where 
\begin{equation}
H_1 = J \sum_{\ell}  \left( 1 + \alpha(u_{\ell} - u_{\ell + 1})\right){\bf S}_{\ell} \cdot {\bf S}_{\ell+1} 
\end{equation}
and
\begin{equation}\label{Eq:3}
H_2 = \sum_{\ell} \left(\frac{P_{\ell}^2}{2M} + \frac{K}{2}(u_{\ell} - u_{\ell + 1})^2\right).
\end{equation}
$u_{\ell}$ and $P_{\ell}$ are the conjugate displacement and momentum operators for the $\ell$th site, and ${\bf S}_{\ell}$ is the Pauli spin operator for that site. 

Defining the phonon frequency $\omega_0 = \sqrt{K/M}$, the lattice degrees of freedom are quantized for dispersive phonons by introducing the phonon creation and annihilation operators,
\begin{equation}\label{}
    b_{\ell}^{\dagger} = \sqrt{\frac{M\omega}{2\hbar}}u_{\ell} -
    i\sqrt{\frac{1}{2M \hbar \omega}}P_{\ell}
\end{equation}
and
\begin{equation}\label{}
   b_{\ell} = \sqrt{\frac{M\omega}{2\hbar}}u_{\ell} +
    i\sqrt{\frac{1}{2M \hbar \omega}}P_{\ell},
\end{equation}
respectively, where $\omega = \sqrt{2}\omega_0$\cite{caron97}.

We solve Eq.\ (\ref{Eq:1})  by the density matrix renormalization group (DMRG) method\cite{caron97,jeckelmann,zhang,barford02} for chains of up to $150$ sites. We use $5$ oscillator levels per site, with typically $200$ block states  and $100,000$  superblock states. One finite lattice sweep is performed at the target chain size. The convergence indicators are shown in Tables \ref{Ta:1} and \ref{Ta:2}. For linear chains with open boundary conditions we maintain constant chains lengths by fixing the position of the end sites.

\begin{table}[tp]
\caption{The ground state energy (in units of $J$) of the Heisenberg spin-Peierls model as a function of the number of sites, $N$. $\omega_0 = J/\sqrt{2}$ and $g=0.1$.}
\begin{center}
\begin{tabular}{ccccccc}
\hline
\hline
  & \multicolumn{6}{c} \textrm{Number of oscillator levels per site} \\

$N$      &  $1$ & $2$  & $3$ & $4$ & $5$ & $6$ \\
\hline
$16$  & $-7.142$ & $-7.330$ & $-7.458$ & $-7.502$ & $-7.518$  & $-7.523$ \\
$28$  & $-12.437$ & $-13.031$ & $-13.286$ & $-13.378$ & $-13.415$  & $-13.430$ \\
$40$ & $-17.746$ & $-18.733$ & $-19.116$ & $-19.256$ & $-19.314$ &  $-19.339$ \\
\hline
\hline
\end{tabular}
\normalsize
\end{center}
\label{Ta:1}
\end{table}

\begin{table}[tp]
\caption{The ground state energy, $E$, (in units of $J$) of the Heisenberg spin-Peierls model as a function of the density matrix eigenvalue product cutoff, $\epsilon$, the number of system block states, $m$, and the superblock Hilbert space size, SBHSS, for a $40$-site chain with $5$ oscillator levels per site. $\omega_0 = J/\sqrt{2}$ and $g=0.1$.}
\begin{center}
\begin{tabular}{cccc}
\hline
\hline
$\epsilon$  & $m$ & SBHSS & $E$  \\
\hline
$10^{-10}$ & $204$ & $35050$ & $-19.311$ \\
$10^{-11}$ & $208$ & $65450$ & $-19.313$ \\
$10^{-12}$ & $104$ & $84450$ & $-19.314$ \\
$10^{-12}$ & $210$ & $106950$ & $-19.314$ \\
\hline
\hline
\end{tabular}
\normalsize
\end{center}
\label{Ta:2}
\end{table}

We first investigate the XY spin-Peierls model for arbitrary phonon frequency and electron-phonon interaction on linear chains with open boundary conditions. For this model,
\begin{equation}
{\bf S}_{\ell} \cdot {\bf S}_{\ell+1} \rightarrow \frac{1}{2}\left({S}^+_{\ell} {S}^-_{\ell+1} + {S}^-_{\ell} {S}^+_{\ell+1}\right).
\end{equation}
Using the Jordon-Wigner transformation the XY model on an open chain is  equivalent to the non-interacting spinless model, where
\begin{eqnarray}\label{Eq:7}
    H_1 \rightarrow  t \sum_{\ell} \left(1 + \alpha({u}_{\ell} - u_{\ell + 1})\right)\left({c}^{\dagger}_{\ell}  {c}_{\ell+1} + \textrm{H.C.}\right)
\end{eqnarray}
and $t \Leftrightarrow J/2$. Following Fradkin and Hirsch\cite{fradkin}, we introduce the dimensionless electron-phonon interaction, $g = \sqrt{\alpha^2t/K} \equiv \sqrt{\lambda\pi/2}$ (where $\lambda$ is the usual definition of the electron-phonon interaction\cite{book}).

The Peierls broken-symmetry ground state is characterized by a non-zero value of the staggered phonon order parameter, defined by
\begin{equation}\label{}
    m(N) = \frac{1}{N}\sum_{\ell}\langle B_{\ell+1}-B_{\ell}\rangle(-1)^{\ell},
\end{equation}
where $B_{\ell} = (b^{\dagger}_{\ell} + b_{\ell})/2$ is the dimensionless displacement of the $\ell$th site and $N$ is the number of sites. To determine whether this order parameter vanishes as a function of the electron-phonon interaction in the asymptotic limit it is necessary to perform a finite size scaling analysis. Fradkin and Hirsch\cite{fradkin} suggested that the scaling relation
\begin{equation}\label{}
    m(N) = \frac{1}{Ng}F(Nm({\infty}))
\end{equation}
should apply. Thus, curves of $Nm(N)$ versus $g$ are expected to coincide at the critical value of $g$ at which $m({\infty})$ vanishes.

\begin{figure}[tb]
\begin{center}
\includegraphics[scale=0.50]{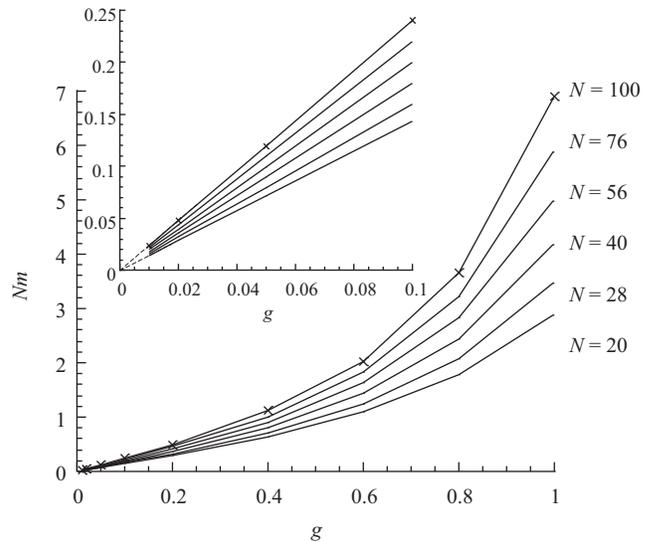}
\end{center}
\caption{The phonon order parameter versus the electron-phonon interaction for the spinless fermion-Peierls model. $\omega_0 = t$. The inset shows a linear extrapolation of $Nm$ versus $g$ to the origin. (The crosses indicate the evaluated points.) The calculations were performed on chains with fixed, open boundary conditions.} \label{Fi:1a}
\end{figure}

\begin{figure}[tb]
\begin{center}
\includegraphics[scale=0.50]{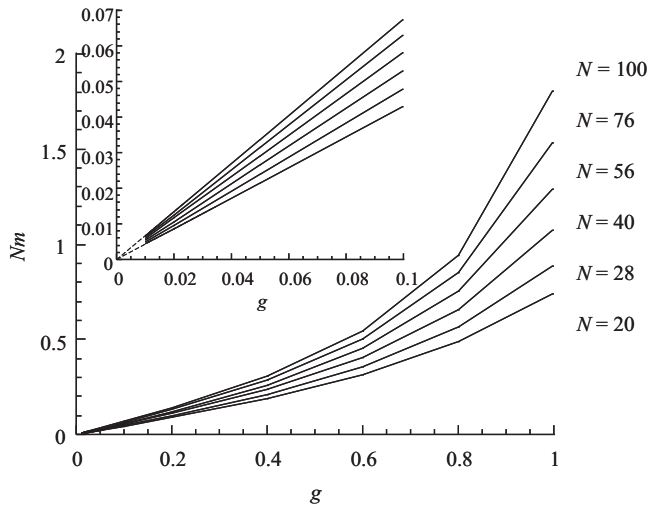}
\end{center}
\caption{As for Fig.\ (\ref{Fi:1a}) with $\omega_0 = 10t$.} \label{Fi:1b}
\end{figure}

Figs (\ref{Fi:1a}) and (\ref{Fi:1b}) show $Nm$ versus $g$ for the spinless fermion-Peierls model (Eqns (\ref{Eq:1}), (\ref{Eq:3}) and (\ref{Eq:7})) for values of $\omega_0 = t$ and $\omega_0 = 10t$, respectively. It is clear in both cases that for small values of $g$ $m$ is linearly proportional to $g$ for all chain lengths. Furthermore, the curves extrapolate linearly to the origin indicating a zero value for the critical $g$. Notice that since the distortion of the $\ell$th bond from its uniform value, $\delta a_{\ell}$, is proportional $\sqrt{\omega_0} g \langle B_{\ell + 1} - B_{\ell} \rangle$, the average staggered bond dimerization is $\propto g^2$ or $\propto \lambda$ for $g \ll 1$.

\begin{figure}[tb]
\begin{center}
\includegraphics[scale=0.50]{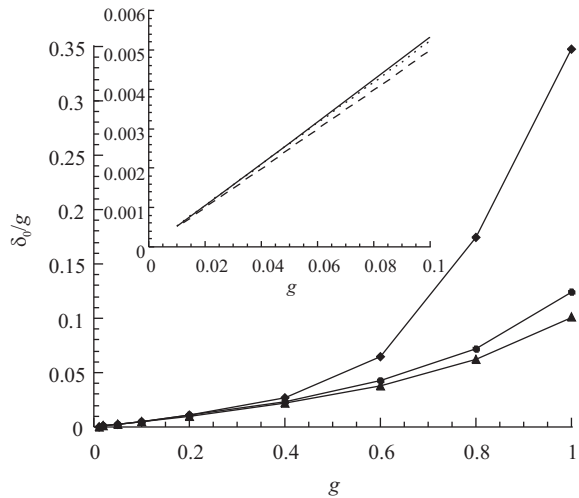}
\end{center}
\caption{The scaled bond dimerization for a $40$-site chain  in the adiabatic limit (diamonds and full line), $\omega_0 = t$ (circles and dotted line), and $\omega_0 = 10t$ (triangles and dashed line).} \label{Fi:2}
\end{figure}

Since the spinless fermion model has an exact solution the order parameter of the spinless fermion-Peierls model is readily obtainable in the adiabatic limit, where the lattice displacements are treated classically. In order to ensure constant chain lengths the classical model is defined as,
\begin{eqnarray}\label{Eq:10}
    H && =   t \sum_{\ell} \left(1 + \alpha({u}_{\ell} - u_{\ell + 1})\right)\left({c}^{\dagger}_{\ell}  {c}_{\ell+1} + \textrm{H.C.}\right)
\nonumber
\\
&& + \sum_{\ell} \left(\frac{K}{2}(u_{\ell} - u_{\ell + 1})^2 + \Gamma(u_{\ell} - u_{\ell + 1}) \right),
\end{eqnarray}
where $\Gamma$ is self-consistently chosen so that $\sum_{\ell}(u_{\ell} - u_{\ell + 1})=0$. The ground state dimerization  of Eq.\ (\ref{Eq:10}) for linear chains is found iteratively  using the Hellmann-Feynman procedure\cite{book}.
We define the normalized, staggered bond dimerization, $\delta_0$, as
\begin{equation} 
\delta_0 = \frac{1}{N}\sum_{\ell} \frac{\delta a_{\ell}}{a}(-1)^{\ell},
\end{equation} 
where $a$ is the average bond length.
Fig.\ (\ref{Fi:2}) shows $\delta_0$ (scaled by $g$) of a $40$-site chain versus $g$ for the two quantum cases considered previously, as well as for the adiabatic limit. Evidently as $g \rightarrow 0$ $\delta_0/g$ tends linearly $\rightarrow 0$. Thus, $\delta_0 \propto g^2 \propto \lambda$. Moreover, the phonon order parameter decreases as a function of $\omega_0$. As $g$ increases the deviation between the quantum and classical predictions increase.

\begin{figure}[tb]
\begin{center}
\includegraphics[scale=0.50]{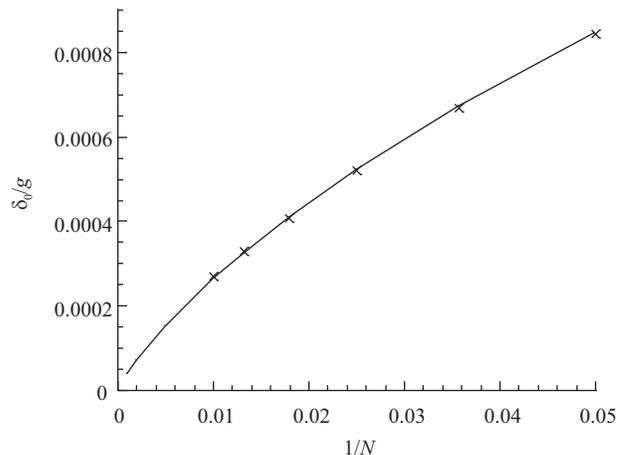}
\end{center}
\caption{The scaled bond dimerization versus inverse chain length in the adiabatic limit. Crosses are the quantum result with $\omega_0 = t$. $g=0.01$.} \label{Fi:3}
\end{figure}

In the adiabatic and asymptotic limits the  bond dimerization satisfies
\begin{equation}\label{Eq:11}
 \delta_0 = 4\exp\left(-\left[1 +\frac{1}{n\lambda}\right]\right)
\end{equation} 
for $\lambda \ll 1$. Here, $n=1$ or $2$ for spinless or spinfull fermions, respectively. However, there are significant finite size corrections to this prediction as $\lambda \rightarrow 0$. Fig.\ (\ref{Fi:3}) shows the bond dimerization versus $1/N$. Also shown are the quantum results for chains of up to $100$ sites. For this small electron-phonon interaction there is excellent agreement between the quantum and classical predictions.

\begin{figure}[tb]
\begin{center}
\includegraphics[scale=0.50]{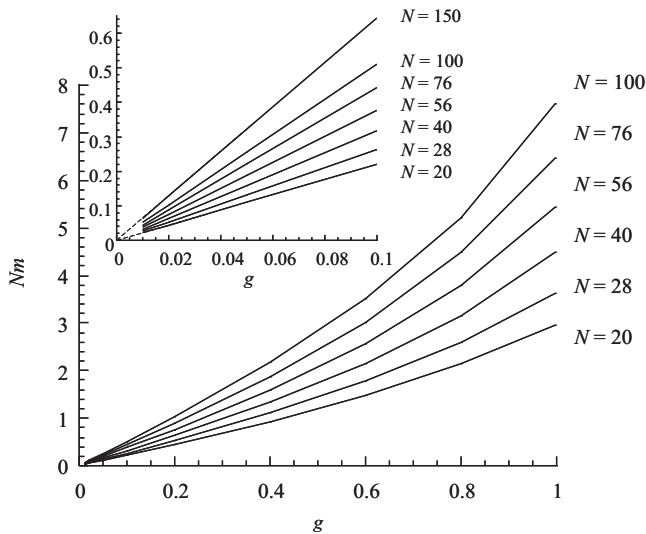}
\end{center}
\caption{The phonon order parameter versus the electron-phonon interaction for the Heisenberg-Peierls model. $\omega_0 = J/\sqrt{2}$. The inset shows a linear extrapolation of $Nm$ versus $g$ to the origin.} \label{Fi:4}
\end{figure}

We now turn to  a discussion of the Heisenberg spin-Peierls model\cite{foot}.
Now the electron-phonon interaction $g = \sqrt{\alpha^2J/K}$.
Fig.\ (\ref{Fi:4}) shows the phonon order parameter versus $g$ for $\omega_0 = J/\sqrt{2}$. The results are consistent with those described earlier, namely $m \propto g$ as $g \rightarrow 0$, indicating a zero critical value of $g$. (Qualitatively similar results are also found for $\omega_0 = 10J/\sqrt{2}$.)

\begin{figure}[tb]
\begin{center}
\includegraphics[scale=0.50]{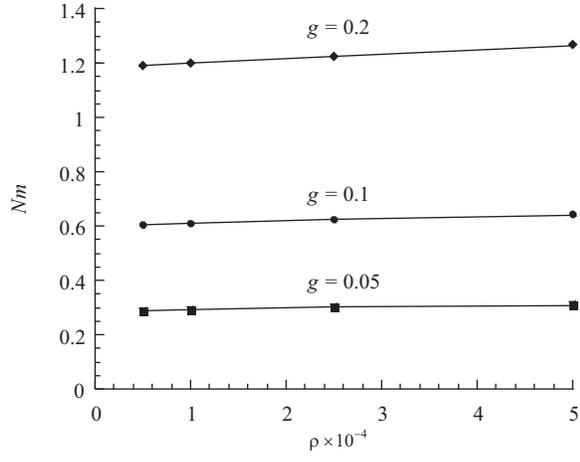}
\end{center}
\caption{The phonon order parameter versus the symmetry breaking parameter $\rho$ for a $100$-site cyclic chain. $\omega_0 = J/\sqrt{2}$.} \label{Fi:5}
\end{figure}

The results described so far have been for linear chains with open boundary conditions. For these chains  the ends break the energetic degeneracy between the otherwise equivalent `A'and `B' phases, which are related by a translation of one repeat unit. In order to check that open boundaries do not introduce any artefacts, we also calculate the phonon order parameter for cyclic chains. In this case it is necessary to explicitly break the degeneracy between the `A' and `B' phases by adding the symmetry breaking term
\begin{equation}\label{}
    \rho \sum_{\ell}  {\bf S}_{\ell} \cdot {\bf S}_{\ell+1} (-1)^{\ell}
\end{equation}
to Eq.\ (1). Fig.\ (\ref{Fi:5}) shows $Nm$ versus $\rho$ for a $100$-site chain for various values of $g$. The results are remarkably consistent with those for open chains. Extrapolating $Nm(\rho)$ to $\rho = 0$ shows that $Nm \propto g$, with values very close to those shown in Fig.\ (\ref{Fi:4}).

In conclusion, we have used the DMRG method to investigate the role of gapless, dispersive quantized phonons on the  Peierls transition in the spin-Peierls model. For both the XY and Heisenberg spin-Peierls model we show that the staggered phonon order scales as $g$ (and the dimerized bond order  scales as $g^2$) as $g \rightarrow 0$. This result is true for both linear and cyclic chains. Thus, we conclude that the Peierls transition occurs at $g=0$. Moreover, for the XY spin-Peierls model we showed that the quantum predictions for the bond order  follow the classical prediction as a function of $1/N$ for small $g$. We therefore conclude that the zero $g$ phase transition is of the mean-field type, Eq.\ (\ref{Eq:11}), with a renormalized effective value of $\lambda$.

\begin{references}
\bibitem{foot} On open chains this model is equivalent to the spinless fermion model with a nearest neighbor interaction $V=J$.

\
\end{references}
\end{document}